# Correlation Between the Energy Shell Structure and Geometry In Metallic Nanoclusters: Quasi-Resonance States, Isotope Effect

Vladimir Kresin

Lawrence Berkeley National Laboratory, Berkeley, CA 94720, USA

Abstract

Metallic nanoclusters displaying electronic shell structure exhibit the special feature of a correlation between their geometry and the number of delocalized electrons . Their shape evolution can be described as a quantum oscillation between quasi-resonant states ( prolate and oblate configurations) whose amplitudes depend upon the degree of shell filling. The picture explains the evolution of absorption spectra and predicts a peculiar isotope effect .

This paper is concerned with the correlation between the electronic structure of metallic nanoclusters and their geometry. This problem is of fundamental value, since the geometry, in turn, affects the electronic structure and spectrum. For this reason it has attracted much interest (see, e.g, [1-7]).

As is known, the electronic states of many metallic nanoclusters form shell structure (see, e.g., the reviews [1-3]). Clusters with filled energy shells ("magic" clusters) are spherical in shape. If the shell is incomplete, there arises an orbital degeneracy which results in the Jahn-Teller (JT) effect, and the shape deforms away from the sphere.

According to calculations in [3-5], as electrons are added to the unoccupied shell, the configuration initially becomes prolate, and then, after more than half a shell is filled, the configuration changes to oblate. Such a conclusion appears convincing, since the calculations yield the minimum-energy configurations. However (as will be demonstrated below) the actual situation is more complicated. Indeed, it is difficult to envision such an abrupt transformation (from prolate to oblate) of a rather large system driven by the addition of just one or two electrons (e.g., $Na_{27}$ → $Na_{28}$, or for ions $Na_{28}^+$ → $Na_{29}^+$). In addition, the prolate and oblate configurations are still characterized by an orbital degeneracy ($\pm m$), where $m$ is the projection of the orbital momentum (although calculations predict that for some clusters this degeneracy is removed by triaxial deformations, see, e.g. [8]).

This paper outlines a different scenario for the evolution of the geometry and electronic structure upon energy shell filling. It will be shown that the proposed picture is in agreement with experimental observations, and suggestions for additional experiments will be presented.

As we know, according to the adiabatic approximation the electronic terms form potential energy surfaces for the nuclear motion, and the ground state of the nuclear configuration corresponds to the minimum of this potential. It is noteworthy that calculations of these terms performed for clusters of different sizes [8] revealed that the term minima for clusters with prolate and oblate configurations, and especially for those that are close to shell half-filling, have very close values of energy ($\Delta E \sim 10^{-2}$ eV).

As mentioned above, if a shell is partially filled, the spherical structure becomes unstable; this instability is caused by the interaction between the electrons and some specific vibrational normal mode: this is the JT effect. For the clusters of interest it is natural to assume that the electronic term(i.e., the potential for the ions) has two equilibrium positions (Fig. 1). In other words, the case is similar to that of a "double-well" potential. Here we have two minima corresponding to two cluster configurations (prolate and oblate) separated by the barrier. As noted above, these configurations are close in energy. Such a case should be approached by methods developed for treating almost degenerate states (see, e.g., [9-10]).

The most convenient approach is the so-called "diabatic" representation introduced in [11] (see also [12]). Then each minimum configuration corresponds to its own potential term, so that the barrier is replaced by the crossing of these terms. As a result, change in configuration can be described as a quantum transition between these terms ( see below, Eqs.(6),(7)). In this representation, the total Hamiltonian can be written in the form:

$$\hat{H} = \hat{H}_{\vec{r}} + \hat{T}_{\vec{R}} + \Delta V \qquad (1)$$

where $\hat{H}_{\vec{r}} = \hat{T}_{\vec{r}} + V(\vec{r},\vec{R}), \Delta V = \tilde{V} - V$, and $\tilde{V}$ corresponds to the crossing of two separate terms (see Fig.1). $\hat{T}_{\vec{r}}$ and $\hat{T}_{\vec{R}}$ are kinetic energy operators for electrons and ions, and $V(\vec{r},\vec{R})$ is the total potential energy; $\vec{r},\vec{R}$ denote the sets of electronic and ionic coordinates. It is fundamentally important that the operator $\hat{H}_{\vec{r}}$, which is diagonal in the usual adiabatic picture, has non-diagonal terms in the "diabatic" representation.

One can seek the total wave function in the form:

$$\Psi(\vec{r},\vec{R}) = a\varphi_a(\vec{r},\vec{R}) + b\varphi_b(\vec{r},\vec{R})$$
*where* (2).
$$\varphi_{a(b)}(\vec{r},\vec{R}) = \tilde{\psi}_{a(b)}(\vec{r},\vec{R})\tilde{\phi}_{a(b)}(\vec{R})$$

*a* and *b* correspond to the prolate and oblate configurations; $\tilde{\psi}_{a(b)}(\vec{r},\vec{R})$ and $\tilde{\phi}_{a(b)}(\vec{R})$ are the electronic and nuclear wave functions in the diabatic representation.

One can see directly from Eq.( 2) that it is impossible to separate electronic and nuclear motions. That is to say, we are dealing with a strongly non-adiabatic phenomenon (the JT effect, see, e.g., [13]).

Substituting the expression (2) into the Schrödinger equation, $\hat{H}\Psi(\vec{r},\vec{R}) = E\Psi(\vec{r},\vec{R})$ we arrive, after some manipulations, at the following secular equation:

$$\text{Det } |A_{ab}|=0, \quad A_{ii}=E-\varepsilon_i \ ; \quad A_{ab}=-\varepsilon_{ab} \ ; \tag{3}$$

$$\varepsilon_i = \int d\vec{R}\tilde{\phi}_i \left[\hat{T}_{\vec{R}} + \hat{H}_{\vec{r};ii}(\vec{R})\right]\tilde{\phi}_i$$
$$\varepsilon_{ab} = \int d\vec{R}\tilde{\phi}_a \left[\hat{T}_{\vec{R}} + \hat{H}_{\vec{r};ab}(\vec{R})\right]\tilde{\phi}_b \tag{3'}$$
$$\hat{H}_{\vec{r};ab}(\vec{R}) = \int \tilde{\psi}_b^* \hat{H}_{\vec{r}} \tilde{\psi}_a d\vec{r}$$

where $i \equiv \{a,b\}$. We neglect small terms such as $\hat{T}_{ab} = \int d\vec{r} d\vec{R} \tilde{\phi}_b \tilde{\phi}_a \tilde{\psi}_b^* \hat{T}_{\vec{R}} \tilde{\psi}_a$, describing the electron-lattice coupling. As a result, we obtain:

$$E_{1,2} = \frac{1}{2}(\varepsilon_a + \varepsilon_b) \mp \left[\frac{1}{4}(\Delta\varepsilon)^2 + \varepsilon_{ab}\varepsilon_{ba}\right]^{1/2} \tag{4}$$

Here $\Delta\varepsilon = \varepsilon_b - \varepsilon_a$; we assume that $\varepsilon_b > \varepsilon_a$.

Eqs. (3) and (4) describing the energy spectrum of clusters with an incomplete shell are a generalization (to the case of two subsystems, electrons and ions) of the conventional expressions for the splitting of degenerate or near-degenerate levels. As was noted above, it is crucial that in the diabatic representation the electronic part of the Hamiltonian contains non-diagonal terms $\hat{H}_{\vec{r};ik}$ (i≠k). As for the wave functions (2), the amplitudes $a_i$ and $b_i$ can be determined for each term from the conditions $a_i^2/b_i^2 = \varepsilon_{ab}^2/(E_i - \varepsilon_a)^2$ ; $a_i^2 + b_i^2 = 1$. For example, for the term $E_1$ we obtain:

$$a_1^2 = \varepsilon_{ab}^2/R \ ; b_1^2 = \{[(\Delta\varepsilon/2)^2 + \varepsilon_{ab}^2]^{1/2} - (\Delta\varepsilon/2)\}^2/R$$
$$R = \{[(\Delta\varepsilon/2)^2 + \varepsilon_{ab}^2]^{1/2} - (\Delta\varepsilon/2)\}^2 + \varepsilon_{ab}^2 \tag{5}$$

In a similar way one can evaluate the coefficients $a_2$ and $b_2$. If $\varepsilon_b \gg \varepsilon_a$, then $\Psi_1 \approx \varphi_a$; $\Psi_2 \approx \varphi_b$.

The electronic state described by Eqs.(2) and (5) might be roughly viewed as a triaxial deformation. However, the picture of a quantum superposition of two configurations described here is very different.

As a shell is filled, there is a continuous decrease in the energy difference, and near a half-filled shell $\Delta\varepsilon \approx 0$. In this case the splitting is determined by the quantity $\varepsilon_{ab}$, and $a_i = b_i = 1/\sqrt{2}$, see Eq.(5): both configurations contribute with equal probabilities.

Cluster behavior also can be studied by means of quantum transition theory. This becomes possible thanks to the presence of non-diagonal terms in the "diabatic" representation. The situation is similar to that with the tunneling effect, which likewise can be treated as a quantum transition [14], [15]. The time-independent operator $\Delta V$ can provide radiationless transitions between the prolate (a) and oblate (b) configurations: $\varphi_a \equiv \varphi^{pr}$, $\varphi_b \equiv \varphi^{obl}$. Let us assume that the wavefunction has the form:

$$\Psi(\vec{r},\vec{R},t) = \alpha(t)\varphi_a(\vec{r},\vec{R}) + \beta(t)\varphi_b(\vec{r},\vec{R}) \tag{6}$$

Imagine that at $t=0$ the cluster is in the prolate configuration;

then α(0)=1, β(0)=0. Using the time-dependent Schrödinger equation, we obtain (see [9,10], cf. [16]):

$$|\beta(t)|^2 = (\varepsilon_{ab}^2/2)\left[(\Delta\varepsilon/2)^2 + \varepsilon_{ab}^2\right]^{-1}\left[1 - \cos 2((\Delta\varepsilon/2)^2 + \varepsilon_{ab}^2)^{1/2} t\right] \qquad (7)$$

We see that the amplitude β(t) - and hence the cluster shape - oscillates between two configurations; the average value of the amplitude β is determined by the relation between $\Delta\varepsilon = \varepsilon_b - \varepsilon_a$ and $\varepsilon_{ab}$; $\varepsilon_a$, $\varepsilon_b$ and $\varepsilon_{ab}$ are determined by Eqs.(3').

It is essential that the spatial scales for the electronic and ionic (vibrational) wave functions are quite different. For the former, the scale is on the order of the bond length, whereas for the latter the scale is on the order of the vibrational amplitude, i.e., much smaller. Based on this fact, one can write (cf. [12],[16])

$$\begin{aligned} \varepsilon_{ab} &= L_0 F \\ L_0 &= \int d\vec{r}\, \tilde{\psi}_b^*(\vec{r}, \vec{R}) \hat{H}_{\vec{r}} \tilde{\psi}_a(\vec{r}, \vec{R})_{|\vec{R}=\vec{R}_0} \\ F &= \int d\vec{R}\, \tilde{\phi}_b(\vec{R}) \tilde{\phi}_a(\vec{R}) \end{aligned} \qquad (8)$$

Therefore, $\varepsilon_{ab}$ can be written as a product of an electronic parameter $L_0 = \hat{H}_{\vec{r};ik}(\vec{R})_{|\vec{R}=\vec{R}_0}$ ($R_0$ corresponds to the region near the crossing point) and a Franck-Condon factor $F$ where the nuclear wave functions $\tilde{\phi}_a$ and $\tilde{\phi}_b$ correspond to different equilibrium configurations. In the harmonic approximation one finds $F \approx \pi A \exp[-1.5(d/A)^2]$. Here $d$ is the distance between the minima, and $A$ is the amplitude of the vibration.

If $\varepsilon_{ab} \gtrsim \Delta\varepsilon$, then $\beta$ strongly depends on $\varepsilon_{ab}$ and consequently on $F$ (if $\varepsilon_{ab} \ll \Delta\varepsilon$, then $\beta \propto \varepsilon_{ab}$ and, therefore, $\beta \propto F$, see Eq. (8)), that is, it depends strongly on the overlap of the vibrational wave functions.

The phenomenon of an oscillation between two structural configurations is similar to that observed in aromatic molecules, e.g., in benzene. This analogy is not accidental. Indeed, both metallic clusters and aromatic molecules are finite systems containing delocalized electrons. The oscillation between prolate and oblate cluster configurations is analogous to oscillations involving the Kekule resonance structures of the benzene molecule (see, e.g., [10],[17]).

Plasmon spectrum.

For magic clusters the photoabsorption spectrum peaks at the frequency of the surface plasmon ( see, e.g., [ 18 ]). For clusters with incomplete shells the spectra are split because of the shape deformation. The photoabsorption cross-section can be written as

$$\sigma(\omega) = \sum_{k=1}^{j} \gamma_k \frac{f_k \omega^2}{\left(\omega^2 - \omega_{0;k}^2\right)^2 + \Gamma_k^2 \omega^2} \tag{9}$$

Here $f_k$ is the oscillator strength, $\Gamma_k$ describes the width of the peak, $\gamma_k = \hbar e^2/mc\Gamma_k$, and $\omega_{0;k}$ are the plasma frequencies: $\omega_{0;1}$ corresponds to collective electron motion along the z-axis (chosen to lie along the symmetry axis), and $\omega_{0;2} = \omega_{0;3}$ are the frequencies along the x and y axes. Note that $\omega_{0;1}^{pr} = \omega_{0;2}^{obl}$.

As has been observed (see, e.g. [19,20]), the spectrum for the cluster with an axial symmetry displays two peaks. For example, for the prolate structure, the lower peak corresponds to the plasmon vibrations along the z-axis $\omega_{01}^{pr}$, whereas the higher peak corresponds to the x and y directions ($\omega_{02}^{pr} = \omega_{03}^{pr}$)

Since the frequencies along the x and y directions are equal, meaning that the corresponding plasmon mode is doubly degenerate, one would expect that the ratio of the effective oscillator strengths would be equal to 1:2 as long as the prolate configuration persists. However, the experimentally observed picture is quite different (see, e.g., Ref.[19]). It is found that the ratio of the amplitudes, $h^{(1)}/h^{(2)}$, which are defined as the areas corresponding to the peaks, actually changes upon shell filling. This ratio increases as the amplitude of the lower peak grows. In the intermediate region, when the shell is close to being half-full, these amplitudes become almost equal to each other.

This dynamics can be easily explained by the picture introduced above, with the state of the cluster described as a superposition of prolate and oblate configurations, see Eqs. (2) and (5). At small shell filling the prolate structure is dominant, and the ratio is close to $h^{(1)}/h^{(2)}=0.5$. On the other hand, close to half-filling the amplitudes of the prolate and oblate configurations are almost equal. Importantly, the oblate and prolate structures yield the same peaks, but with a reverse relation between the oscillator strengths. As a result, near a half-filled shell (e.g., close to $N=27$, that is, e.g. for $Na_{28}^+$), one should expect both peaks to display similar intensities, as observed experimentally.

The imbalance $h^{(1)}/h^{(2)} \neq 0.5$ was explained in some papers (e.g., [6,8 ,21]) by the presence of isomers with prolate and oblate structures. Such a picture could, in principle, explain the above inequality. However, isomers, similarly to triaxial structures, represent *isolated* clusters in specific stationary states. Here, on the other hand, we discuss the quantum superposition of states for an *individual* cluster; these could be called resonant (or quasi-resonant) states. Such a description is more general and rigorous. Isomers represent the limiting case of a resonant structure, corresponding to a negligibly small value of the quantity $\varepsilon_{ab}$: in this limit the lifetime for each configuration [see Eq.(7)] is very large, which corresponds to an almost stationary state.

It appears, that the dynamics of clusters with incomplete shell, and especially the region near half-filling, are better described by the picture and language of a quantum superposition of configurations. Furthermore, this picture predicts a peculiar isotope effect which would be absent in the isomer scenario. This isotope effect will be described in the next section.

<u>Isotope effect</u>

The strong non-adiabaticity reflected in Eqs.(2),(6) brings up the appearance of some peculiar phenomena. Among them is the isotope variation of the absorption spectrum. At first glance this sounds strange, since the absorption spectrum is dominated by the electronic oscillator strength, whereas isotope substitution affects the ionic subsystem. However, the mixing of electronic and nuclear motions caused by non-adibaticity can produce such an effect. More precisely, the isotope effect is caused by the presence of the Franck-Condon (FC) factor in the matrix elements, Eq.(8). Indeed, consider the quantity

$$B = (h^{(1)}/h^{(2)}) - 0.5 \qquad (10)$$

If there is one electron in the unoccupied shell, then the prolate configuration strongly dominates and $B \approx 0$. Adding more electrons increases the contribution of the oblate configuration. The degree of mixing of the two configurations is determined by the quantity $\tilde{\beta}^2$, see Eqs. (6) and (7); $\tilde{\beta}$ is the average value of the amplitude $\beta$. As discussed above, the relation between two peaks in the absorption spectrum is governed by the same phenomenon of mixing. Therefore, $B \propto \tilde{\beta}^2$, that is $B = c\tilde{\beta}^2$, c=const.

For a half-filled shell $h^{(1)} = h^{(2)}$, and $B = 0.5$. Therefore, $c = 1$ (see Eq.(10)) and we obtain $B = \tilde{\beta}^2$. Correspondingly [see Eq.(7)],

$$B = (\varepsilon_{ab}^2/2)\left[(\Delta\varepsilon/2)^2 + \varepsilon_{ab}^2\right]^{-1}. \qquad (11)$$

The parameter $\varepsilon_{ab}$ contains the electronic and the Franck-Condon factors, see Eq.(8). The FC factor strongly (exponentially) depends on the ionic mass, since $A^2 \propto M^{-1}$. The quantity $\Delta\varepsilon$ decreases as the shell is filled, and near half-filling $\Delta\varepsilon \approx 0$. Then $B$ does not depend on $M$ [see Eq.(11)]. The dependence $\Delta\varepsilon(N)$ can be written in the form $\Delta\varepsilon=\varepsilon_0[(N_m'/2)-N^s]^2$; here $\varepsilon_0$ is a parameter, and $N_m'$ is the number of electrons in the specific complete shell (it corresponds to the "magic" number $N_m$, where $N_m$ is a total number of delocalized electrons for the cluster with a complete shell), $N^s$ is a number of electrons in this shell. In other words, the quantity $(N_m'/2)-N^s$ is the deviation from the case of half-filled shell. As a result, the quantity $B$ becomes $B=(v^2/2)[(N_m'/2)-N^s)^2+v^2]^{-1}$, with $v=\varepsilon_{ab}/\varepsilon_0$.

The isotopic dependence can be described by the relation $B \propto M^{-\eta}$, where $\eta$ is the isotope coefficient. Correspondingly, $\eta = -(M/B)(dB/dM)$. After a straightforward calculation, we obtain the following expression for the isotope coefficient:

$$\eta = (B/2v^2)\left[(N_m'/2)-N^s\right]^2 (d/A)^2 \qquad (12)$$

The dependence (12) can be verified experimentally. Note also that the quantity $B$ defined by Eq.(10) can be determined directly from measurements, see, e.g., Ref.[19]; this allows us to evaluate the parameter $v$.

Note that the presence of thermal fluctuations also can lead to transitions between configurations. But being strongly temperature dependent, such channel can be separated by measurements at various temperatures.

Consider, for example, the alkali clusters $M_N$ such as Na , K, Li in the size range 20<*N*<34 (*N*=20 and *N*=34 are "magic" numbers, see ,e.g.,[22]).  Detailed measurements of the spectrum for Na clusters which allow a direct determination of *B* were described in [20], see also [ 19,23 ]. One can expect that K and Li  clusters are characterized by similar parameters.

Based on Eq.(12) we conclude that the largest values of the isotope coefficient correspond to clusters with slightly filled shells, for example, those with *N*=21,…,24.  In contrast, for the clusters with an almost half-filled shell (*N*=27, then *B*=0) the derivative  is small, and the isotopic dependence is weak.

For example, for the cluster $M_{23}$ we obtain the value $\eta \approx 0.95$, for $M_{25}$ we have $\eta \approx 0.7$. These are rather large numbers, and it would be interesting to search for such isotopic dependence experimentally. Probably,  lithium clusters are the best for such an experiments, because Li, unlike Na, exists in two stable isotopic forms $^7$Li and $^6$Li.  In this case, isotope substitution will lead to a noticeable change in the parameter *B* whose value, according to Eq. (10) can be measured directly. In principle, one can study also K clusters which are characterized by stable isotopes $^{39}$K and  $^{41}$K.

For example, for the cluster $K_{23}$ isotope substitution $^{39}K \rightarrow {}^{41}K$ leads to a shift of $\delta B/B \approx 5\%$. Larger shift could be observed for Li clusters. Namely, the substitution $^7Li \rightarrow {}^6Li$ would lead to the shift $\delta B/B \approx 14\%$.

An experiment to detect such a shift would support our concept of a superposition of structures and its evolution with shell filling.

The author is grateful to J. Friedel and V. V. Kresin for interesting discussions. The research was supported by DARPA.

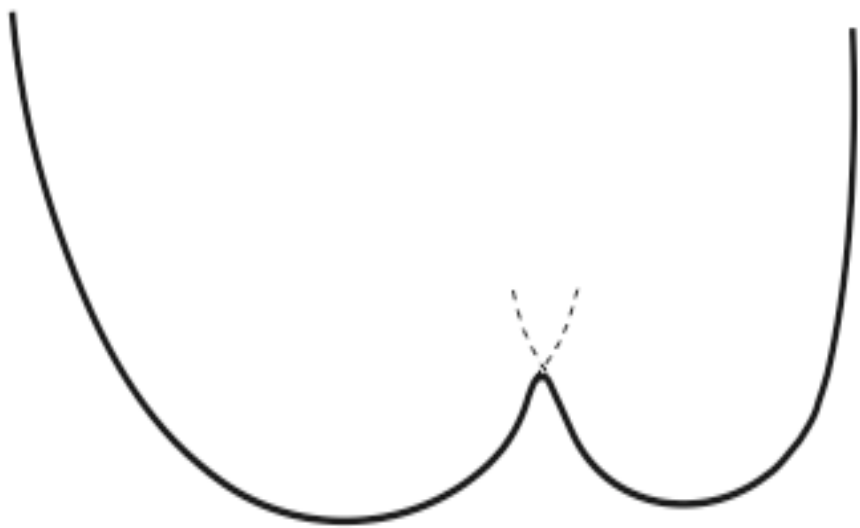

**Fig.1**

Diabatic representation (1D model). Solid line: adiabatic potential energy term; dashed line corresponds to the crossing terms in the diabatic representation.